\documentclass[namedreferences]{solarphysics}

\usepackage[hyperref,optionalrh,solaromanenum]{spr-sola-addons}
\usepackage{graphicx}
\usepackage{color}   
\usepackage{breakurl}

\newcommand{\aap}{Astron.\ Astrophys.}%
\newcommand{\aj}{Astron.\ J.}%
\newcommand{\apj}{Astrophys.\ J.}%
\newcommand{\apjl}{Astrophys.\ J.\ Lett.}%
\newcommand{\apjs}{Astrophys.\ J.\ Supp.}%
\newcommand{\mnras}{Mon.\ Not.\ Roy.\ Astron.\ Soc.}%
\newcommand{\nat}{Nature}%
\newcommand{\procspie}{\it Proc.~SPIE}%
\newcommand{\pasp}{Pub.\ Astron.\ Soc.\ Pacific}%

\begin{document}
\begin{article}

\begin{opening}

\title{Magnetic Evolution and the Disappearance of Sun-like Activity Cycles}

\author[addressref={1,2},corref,email={tmetcalfe@spacescience.org}]{\inits{T.S.~}\fnm{Travis~S.~}\lnm{Metcalfe}}%
\author[addressref={3,4},corref,email={jvansaders@obs.carnegiescience.edu}]{\inits{J.L.~}\fnm{Jennifer~}\lnm{van~Saders}}

\address[id=1]{Space Science Institute, 4750 Walnut Street, Suite 205, Boulder CO 80301 USA}
\address[id=2]{White Dwarf Research Corp., 3265 Foundry Place, Unit 101, Boulder CO 80301 USA}
\address[id=3]{The Observatories of the Carnegie Institution for Science, 813 Santa Barbara Street, Pasadena CA 91101 USA}
\address[id=4]{Dept.\ of Astrophysical Sciences, Princeton University, Princeton NJ 08544 USA}

\runningauthor{Metcalfe \& van Saders}
\runningtitle{Disappearance of Sun-like Cycles}

\begin{abstract} 
After decades of effort, the solar activity cycle is exceptionally well 
characterized but it remains poorly understood. Pioneering work at the 
Mount Wilson Observatory demonstrated that other sun-like stars also show 
regular activity cycles, and suggested two possible relationships between 
the rotation rate and the length of the cycle. Neither of these 
relationships correctly describe the properties of the Sun, a peculiarity 
that demands explanation. Recent discoveries have started to shed light on 
this issue, suggesting that the Sun's rotation rate and magnetic field are 
currently in a transitional phase that occurs in all middle-aged stars. 
Motivated by these developments, we identify the manifestation of this 
magnetic transition in the best available data on stellar cycles. We 
propose a reinterpretation of previously published observations to suggest 
that the solar cycle may be growing longer on stellar evolutionary 
timescales, and that the cycle might disappear sometime in the next 
0.8-2.4~Gyr. Future tests of this hypothesis will come from ground-based 
activity monitoring of Kepler targets that span the magnetic transition, 
and from asteroseismology with the TESS mission to determine precise 
masses and ages for bright stars with known cycles. 
\end{abstract}


\end{opening}

% SECTION 1 %%%%%%%%%%%%%%%%%%%%%%%%%%%%%%%%%%%%%%%%%%%%%%%%%%%%%%%%%%%%%%

\section{Astrophysical Context}\label{sec1} 

The periodic rise and fall in the number of sunspots every 11 years was 
first noted by \cite{Schwabe1844}, and the detailed patterns of spot 
orientation and migration throughout this solar activity cycle have 
subsequently been characterized with exquisite observations spanning many 
decades. Stellar dynamo theory attempts to understand these patterns by 
invoking a combination of convection, differential rotation, and 
meridional circulation to modulate the global magnetic field 
\citep[see][]{Charbonneau2010}. Observations of other sun-like stars are 
necessarily more limited because in most cases we cannot spatially resolve 
spots on their surfaces. However, the solar activity cycle is clearly 
detectable without spatial resolution from observations of the intensity 
of emission in the Ca~{\sc ii}~H (396.8~nm) and K (393.4~nm) spectral 
lines (hereafter Ca\,HK). These lines have long been used as a proxy for 
the strength and filling factor of magnetic field because the emission 
traces the amount of non-radiative heating in the chromosphere 
\citep{Leighton1959}. The most comprehensive spectroscopic survey for 
Ca\,HK variations in sun-like stars was conducted over more than 30 years 
from the Mount Wilson Observatory \citep{Wilson1978, Baliunas1995}, 
yielding the first large sample of stars with measured rotation rates and 
activity variations to help validate stellar dynamo theory.

Initial results from the Mount Wilson sample suggested that both the 
stellar cycle period and the mean activity level depend on the Rossby 
number, the rotation rate normalized by the convective turnover time 
\citep[Ro\,$\equiv P_{\rm rot}/\tau_c$, see][]{Noyes1984}. Cycle periods 
were shortest for the most rapidly rotating young stars, while they were 
longer for older stars with slower rotation. \cite{Brandenburg1998} 
suggested that there were actually two distinct relationships between the 
rotation rate and the length of the cycle, with an upper sequence of stars 
showing a cycle every 300-500 rotations, and a lower sequence of shorter 
cycles requiring fewer than $\sim$100 rotations. At moderate rotation 
rates (10-22 days), some stars exhibited cycles simultaneously on both 
sequences. \cite{BohmVitense2007} interpreted this dual pattern as 
evidence for two stellar dynamos operating in different shear layers, 
possibly at the bottom of the outer convection zone (the tachocline), or 
in the near-surface regions as suggested by helioseismic inversions 
\citep{Thompson1996}.

One of the most perplexing results from the Mount Wilson survey is that 
neither of the stellar-based relationships between the length of the cycle 
and the rotation rate correctly describe the properties of the Sun. With a 
mean cycle period of 11 years and a sidereal rotation period of 25.4 days 
($P_{\rm cyc}/P_{\rm rot}\!\!\sim$160), the Sun falls between the two 
stellar sequences \citep{BohmVitense2007}. Recent work may have identified 
the reason why the solar activity cycle does not fit the pattern 
established by other stars: the Sun's rotation rate and magnetic field may 
be in a transitional phase that occurs in all middle-aged stars 
\citep{vanSaders2016, Metcalfe2016}. In the following section, we review 
the evidence for a magnetic transition and the underlying mechanisms. In 
Section~\ref{sec3}, we examine previously published observations to 
identify the manifestation of this transition in stellar activity cycles, 
and we propose a new scenario for the evolution of the solar cycle. We 
discuss the potential for future observational tests of this hypothesis in 
Section~\ref{sec4}.

% SECTION 2 %%%%%%%%%%%%%%%%%%%%%%%%%%%%%%%%%%%%%%%%%%%%%%%%%%%%%%%%%%%%%%

\section{Magnetic Metamorphosis}\label{sec2}

The idea of using rotation as a diagnostic of stellar age dates back to 
\cite{Skumanich1972}, and a decade of effort has gone into calibrating the 
modern concept of {\it gyrochronology} \citep{Barnes2007}. Although stars 
are formed with a range of initial rotation rates, the stellar winds 
entrained in their magnetic fields lead to angular momentum loss from 
magnetic braking \citep[see][]{Kawaler1988}. The angular momentum loss 
scales strongly with the angular rotation velocity $dJ/dt\propto\omega^3$, 
which forces convergence to a single rotation rate at a given mass after 
roughly 500~Myr in sun-like stars \citep{Pinsonneault1989}. The evidence 
for this scenario relies on studies of rotation in young clusters at 
various ages, and until recently the only calibration point for ages 
beyond $\sim$1~Gyr was from the Sun.

The situation changed after the Kepler space telescope provided new data 
for older clusters and field stars. The initial contributions from Kepler 
included observations of stellar rotation in the 1~Gyr-old cluster 
NGC\,6811 \citep{Meibom2011} and the 2.5~Gyr-old cluster NGC\,6819 
\citep{Meibom2015}, extending the calibration of gyrochronology 
significantly beyond previous work. The first surprises emerged when 
asteroseismic ages became available for Kepler field stars with measured 
rotation periods \citep{Metcalfe2014, Garcia2014}. Initial indications of 
a possible conflict between asteroseismology and gyrochronology were noted 
by \cite{Angus2015}, who found that no single mass-dependent relationship 
between rotation and age could simultaneously describe the cluster and 
field populations. Although they used low-precision asteroseismic ages 
from grid-based modeling \citep{Chaplin2014}, the tension was still 
evident.

% 2.1
\subsection{Breakdown of Magnetic Braking}

The source of disagreement between the age scales from asteroseismology 
and gyrochronology came into focus after \cite{vanSaders2016} scrutinized 
Kepler targets with precise ages from detailed modeling of the individual 
oscillation frequencies \citep{Mathur2012, Metcalfe2012, Metcalfe2014}. 
They confirmed the existence of a population of field stars rotating more 
quickly than expected from gyrochronology. They discovered that the 
anomalous rotation became significant near the solar age for G-type stars, 
but it appeared $\sim$2-3~Gyr for hotter F-type stars and $\sim$6-7~Gyr 
for cooler K-type stars. This dependence on spectral type suggested a 
connection to the Rossby number, because cooler stars have deeper 
convection zones with longer turnover times. They postulated that magnetic 
braking may operate with a dramatically reduced efficiency beyond a 
critical Rossby number, and they reproduced the observations with models 
that eliminated angular momentum loss beyond Ro\,$\sim$\,2. This value is 
derived from a model-dependent estimate of the convective turnover time 
one pressure scale-height above the base of the outer convection zone. 
Although the specific value obtained by \cite{vanSaders2016} depends on 
mixing-length theory, the observed trend for stars of various masses and 
ages is robust.

% FIGURE 1 --------------------------------------------------------------- 
 \begin{figure*}[t] 
 \centerline{\includegraphics[angle=270,width=\textwidth]{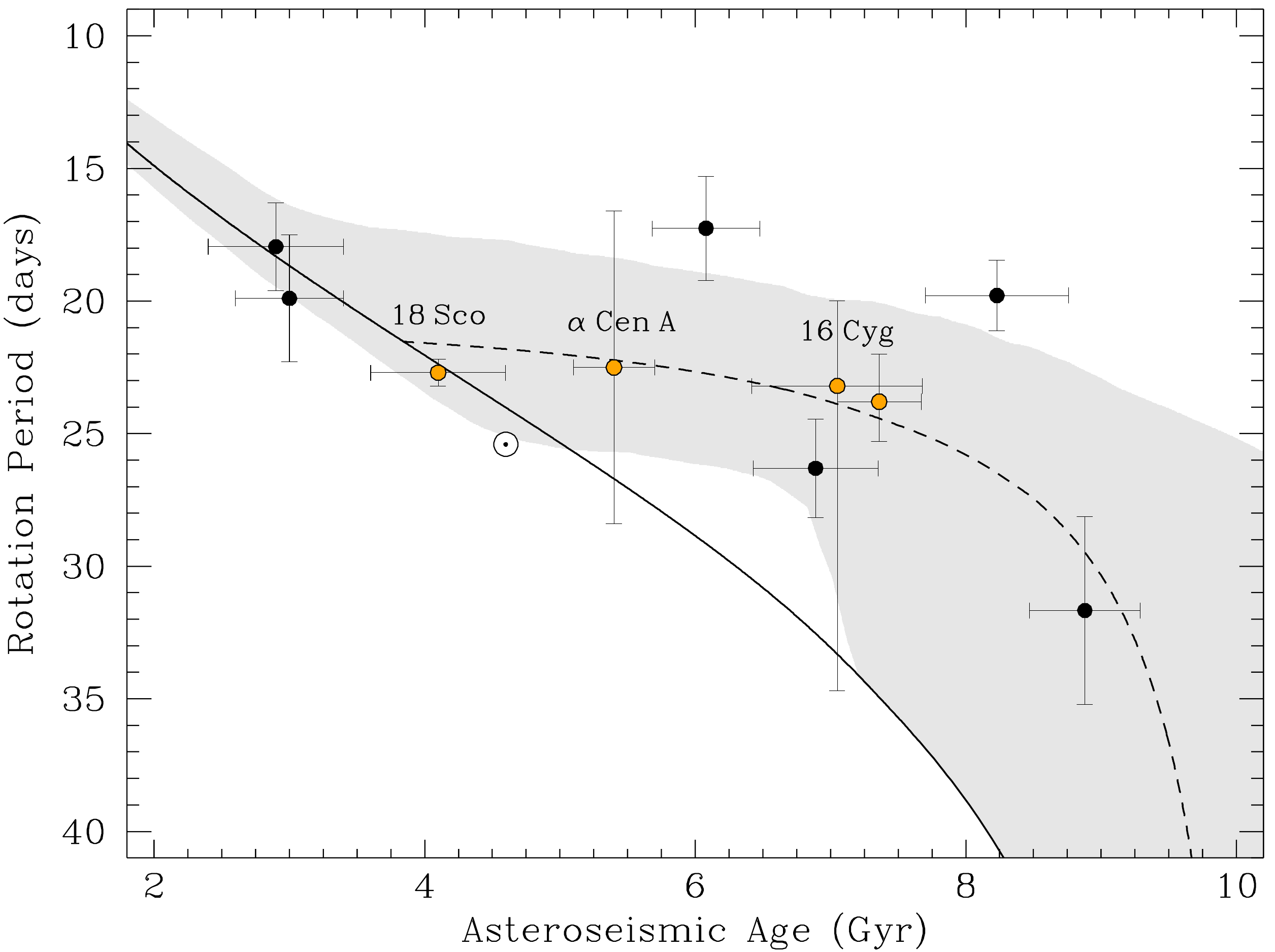}} 
\caption{Stellar evidence for the shutdown of magnetic braking in sun-like 
stars. The solid line shows a standard rotational evolution model, which 
is calibrated using young star clusters and the Sun. The dashed line shows 
the modified model of \cite{vanSaders2016} for solar metallicity and a 
zero age main-sequence (ZAMS) effective temperature of 5750~K, which 
eliminates angular momentum loss beyond a critical Rossby number 
(Ro$\sim$2), determined from a fit to Kepler field stars with 
asteroseismic ages (black points and 16\,Cyg). The shaded region 
represents the expected dispersion due to the range of masses and 
metallicities within the field star sample, encompassing ZAMS effective 
temperatures between 5600-5900~K. A few well-characterized solar analogs 
are shown with yellow points.}\label{fig1}
 \end{figure*}
%-------------------------------------------------------------------------

The anomalous rotation discovered by \cite{vanSaders2016} is illustrated 
for sun-like stars in Figure~\ref{fig1}. A standard rotational evolution 
model (solid line) and the modified model that eliminates angular momentum 
loss beyond a critical Rossby number (dashed line) are from the original 
paper, which also used hotter and cooler stars to constrain the fit. Note 
that the solar age and rotation rate (marked with a $\odot$ symbol) were 
used to calibrate the standard model beyond the 0.5-2.5~Gyr age range of 
clusters. Asteroseismic ages for the Kepler sample (black points and 
16\,Cyg) have been updated with values from \cite{Creevey2017}. The shaded 
region represents the expected dispersion due to the range of masses and 
metallicities within the sample (e.g., the two high points are lower 
metallicity stars, giving them thinner convection zones that reach the 
critical Rossby number at faster rotation rates). The asteroseismic 
rotation rates and ages for a $\sim$3~Gyr-old solar analog binary system 
\citep{White2016} have been overplotted, validating asteroseismic rotation 
measurements and the age scale for sun-like Kepler stars. A few 
well-characterized solar analogs are shown with yellow points, including 
18\,Sco \citep{Petit2008, Li2012, Mittag2016}, $\alpha$\,Cen\,A 
\citep{Bazot2007, Bazot2012}, and 16\,Cyg\,A\,\&\,B \citep{Davies2015, 
Creevey2017}. Although some uncertainties remain for 18\,Sco and 
$\alpha$\,Cen\,A, these bright stars appear to follow the same pattern of 
anomalous rotation observed in the Kepler sample.

\cite{vanSaders2016} suggested that magnetic braking might become less 
efficient in older stars from a concentration of the field into smaller 
spatial scales. \cite{Reville2015} demonstrated that the dipole component 
of the global field is responsible for most of the angular momentum loss 
due to the magnetized stellar wind \citep[see also][]{Garraffo2016}. The 
Alfv\'en radius is greater for the larger scale components of the field, 
and because both the open flux and the effective lever-arm increase with 
increasing Alfven radius, low-order fields consequently shed more angular 
momentum. The inverse of this process may be responsible for the onset of 
efficient magnetic braking in very young stars \citep{Brown2014}.

% 2.2
\subsection{Triggering the Magnetic Transition}

\cite{Metcalfe2016} identified a magnetic counterpart to the rotational 
transition discovered by \cite{vanSaders2016}. They compiled published 
Ca\,HK measurements for the Kepler sample and compared them to a selection 
of sun-like stars from the Mount Wilson survey \citep{Baliunas1996, 
Donahue1996}. Such a comparison requires the Ca\,HK measurements to be 
converted to a chromospheric activity scale ($\log R'_{\rm HK}$) that 
accounts for the bolometric flux of different spectral types.

\begin{figure*}[t] 
 \centerline{\includegraphics[angle=270,width=\textwidth]{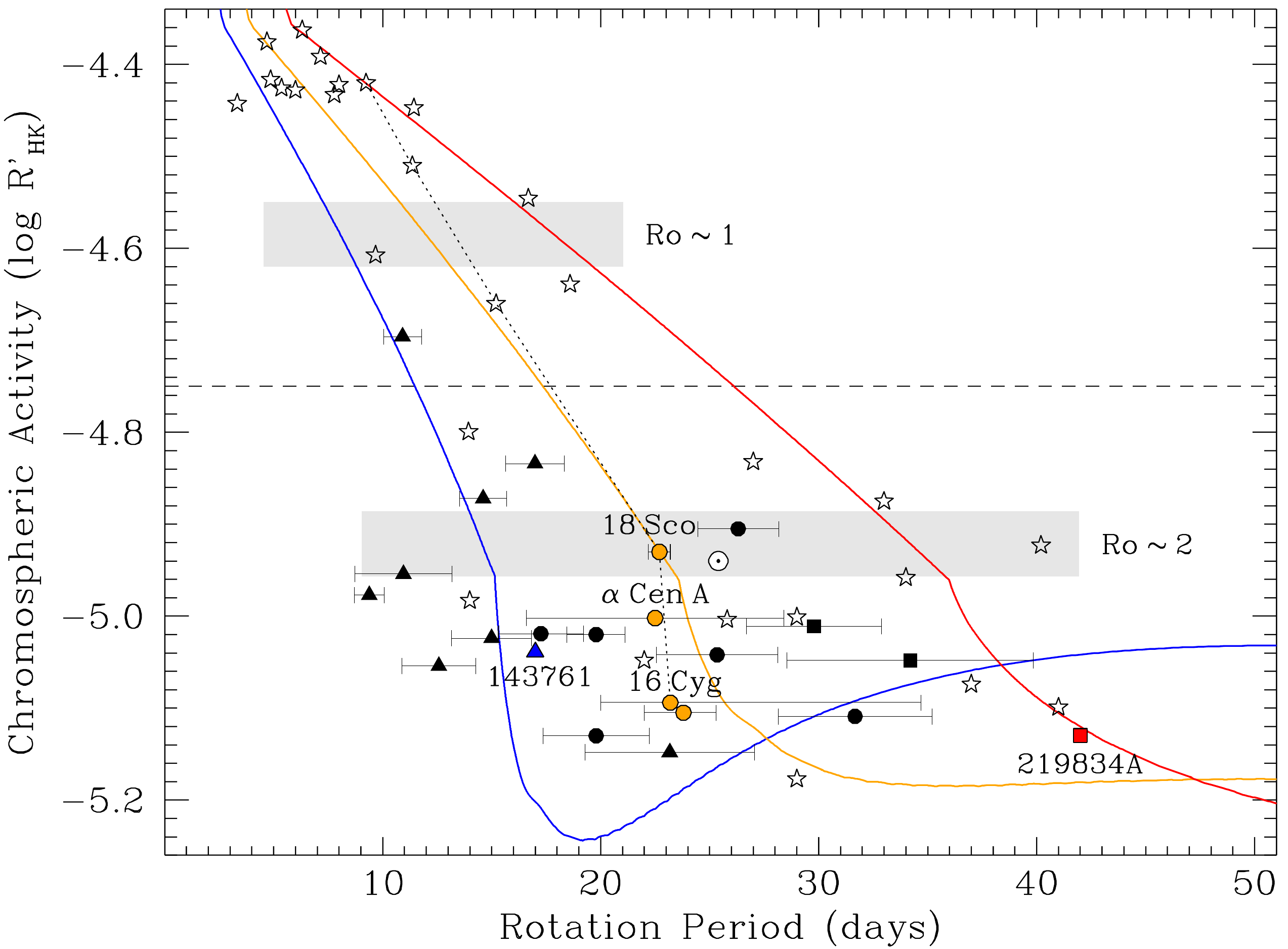}} 
\caption{Relationship between rotation and chromospheric activity in field 
dwarfs and subgiants. The asteroseismic sample from Kepler is shown with 
black points, and a selection of targets from the Mount Wilson survey are 
shown as star symbols. The solar analogs discussed in \cite{Metcalfe2016} 
are connected with a dotted line, crossing the Vaughan-Preston gap (dashed 
line) before reaching the critical Rossby number (Ro$\sim$2, shaded 
region) where magnetic braking becomes less efficient. The solar analogs 
from Figure~\ref{fig1} are again shown with yellow points, the F-type star 
HD\,143761 is shown as a blue triangle, and the K-type star HD\,219834A is 
shown as a red square.}\label{fig2}
 \end{figure*}

The relationship between chromospheric activity and rotation is 
illustrated in Figure~\ref{fig2}. The Kepler targets are plotted by 
spectral type, including F-type (triangles), G-type (circles), and K-type 
stars (squares), while the Mount Wilson targets are shown as star symbols. 
Several rotational evolution models from \cite{vanSaders2016} are shown, 
converted from Rossby number to chromospheric activity using the 
rotation-activity relation of \cite{Mamajek2008}. The activity levels that 
correspond to key Rossby numbers are shown as shaded regions on either 
side of the Vaughan-Preston gap \citep[dashed line;][]{Vaughan1980}. The 
dotted line connects some well-characterized solar analogs, including the 
same stars shown with yellow points in Figure~\ref{fig1}.

% FIGURE 2 --------------------------------------------------------------- 
%-------------------------------------------------------------------------

The magnetic evolution of sun-like stars appears to change dramatically 
when they reach the critical Rossby number (Ro$\sim$2) identified by 
\cite{vanSaders2016}. The shutdown of magnetic braking near the activity 
level of 18\,Sco \citep[$\log R'_{\rm HK}$=$-4.93$;][]{Hall2007a} keeps 
the rotation rate nearly constant as the activity level continues to 
decrease with age toward $\alpha$\,Cen\,A \citep[$\log R'_{\rm 
HK}$=$-5.00$;][]{Henry1996} and 16~Cyg \citep[$\log R'_{\rm 
HK}$=$-5.09$;][]{Wright2004}. A similar transition occurs at faster 
rotation rates for hotter stars like HD\,143761 (blue triangle), and at 
slower rotation rates for cooler stars like HD\,219834A (red square). The 
influence of this magnetic transition on stellar activity cycles is 
described in Section~\ref{sec3}.

\cite{Metcalfe2016} proposed that a change in the character of 
differential rotation is the mechanism that ultimately disrupts the 
large-scale organization of magnetic fields in sun-like stars. The process 
begins at Ro\,$\sim$\,1, where the rotation period becomes comparable to 
the convective turnover time. Differential rotation is an emergent 
property of turbulent convection in the presence of Coriolis forces, and 
\cite{Gastine2014} showed that many global convection simulations exhibit 
a transition from solar-like to anti-solar differential rotation near 
Ro\,$\sim$\,1 \citep[see also][]{Brun2017}. The Vaughan-Preston gap can 
then be interpreted as a signature of rapid magnetic evolution triggered 
by a shift in the character of differential rotation. \cite{Pace2009} used 
activity measurements of stars in several open clusters to constrain the 
age of F-type stars crossing the gap to be between 1.2 and 1.4~Gyr. The 
two most active F-type stars in the Kepler sample have ages of 0.94 and 
1.64~Gyr and fall on opposite sides of the gap, again validating the 
asteroseismic age scale.

Emerging from the rapid magnetic evolution across the Vaughan-Preston gap, 
stars reach the Ro\,$\sim$\,2 threshold where magnetic braking operates 
with a dramatically reduced efficiency, possibly due to a shift in 
magnetic topology. The rotation period then evolves as the star undergoes 
slow expansion and changes its moment of inertia as it ages. At the same 
time, the activity level decreases with effective temperature as the star 
expands and mechanical energy from convection largely replaces magnetic 
energy driven by rotation as the dominant source of chromospheric heating 
\citep{BohmVitense2007}.

% SECTION 3 %%%%%%%%%%%%%%%%%%%%%%%%%%%%%%%%%%%%%%%%%%%%%%%%%%%%%%%%%%%%%%

\section{Manifestation in Stellar Activity Cycles}\label{sec3}

\begin{figure*}[t] 
 \centerline{\includegraphics[angle=270,width=\textwidth]{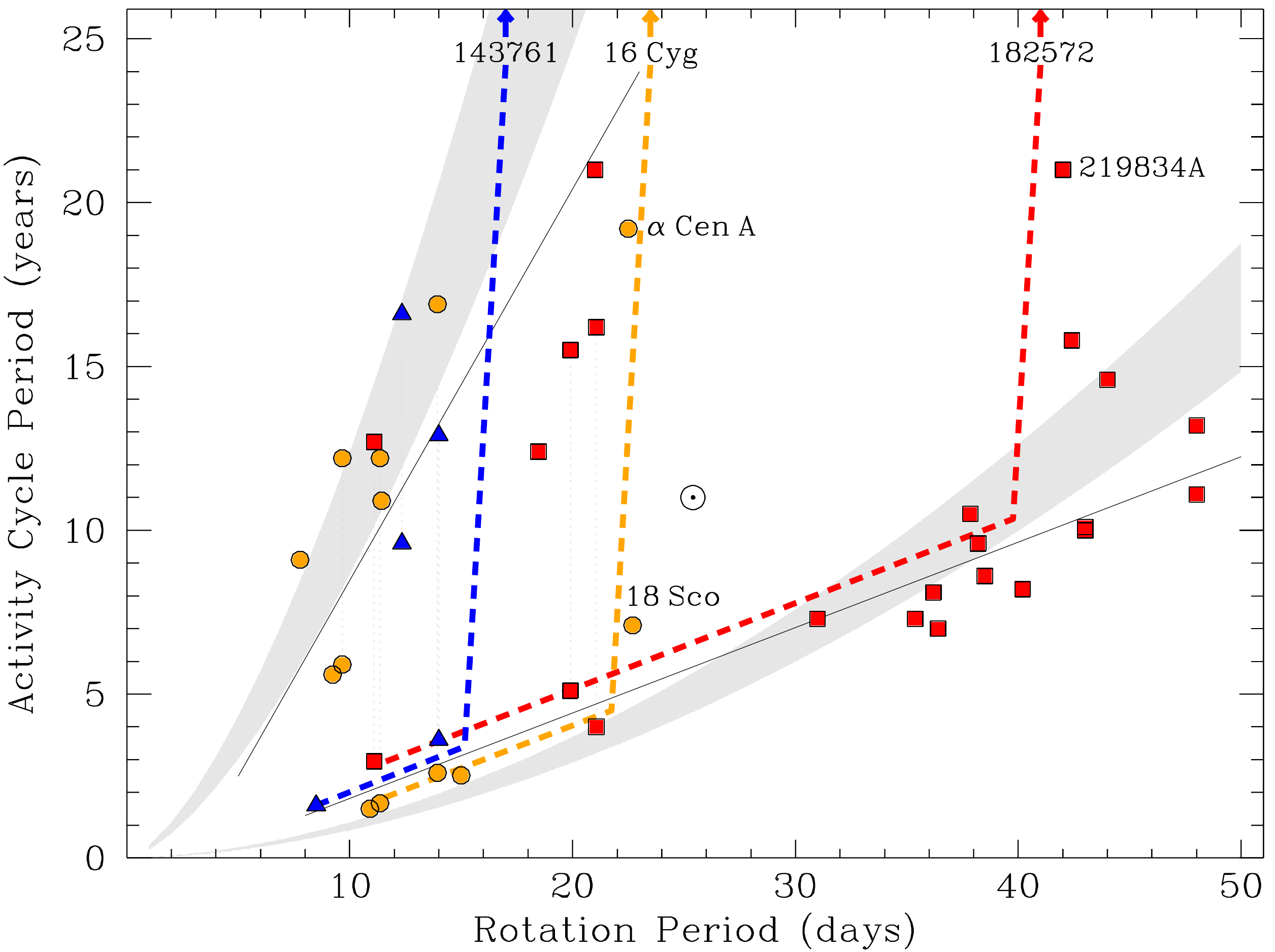}} 
\caption{Updated version of a diagram originally published by 
\cite{BohmVitense2007}, showing two different relationships between 
rotation rate and the length of the activity cycle. Cycles operating 
simultaneously in the same star are connected with a vertical dotted line. 
Points are colored by spectral type, indicating F-type (blue triangles), 
G-type (yellow circles), and K-type stars (red squares). Schematic 
evolutionary tracks are shown as dashed lines, leading to stars that 
appear to have completed the magnetic transition. Shaded regions indicate 
the fits of \cite{Brandenburg1998} using a range of convective turnover 
times appropriate for G-type stars ($\tau$=7-14~days) on the upper 
sequence and for K-type stars ($\tau$=17-23~days) on the lower 
sequence.}\label{fig3}
 \end{figure*}

The new picture of rotational and magnetic evolution provides a framework 
for understanding some observational features of stellar activity cycles 
that have until now been mysterious. An updated version of a diagram 
published in \cite{BohmVitense2007} is shown in Figure~\ref{fig3}, using 
data from \cite{Brandenburg1998}. More recent data have been added from 
\cite{Hall2007b}, \cite{Bazot2007}, \cite{Petit2008}, \cite{DeWarf2010}, 
\cite{Metcalfe2010, Metcalfe2013}, \cite{Ayres2014}, \cite{Egeland2015}, 
and \cite{Salabert2016}. We do not include marginal detections of stellar 
cycles that may obscure the relationships suggested by the best available 
data \citep[e.g., see][]{Egeland2017}.

The stellar sequence along the bottom of Figure~\ref{fig3} has three 
distinct regimes. For faster rotators ($P_{\rm rot}<22$~days), this 
sequence is dominated by short cycles for stars that also show longer 
cycles on the upper sequence (vertical dotted lines). Many of the Mount 
Wilson targets in this regime appeared to have ``chaotic variability'' in 
their chromospheric activity. This may be due to the ubiquity of short 
period cycles on the lower sequence, combined with seasonal data gaps that 
failed to sample these timescales adequately. F-type stars are expected to 
begin the magnetic transition at rotation periods $\sim$15~days, but there 
are very few hot stars with well determined cycles. The oldest cycling 
F-type star in the Mount Wilson sample is HD\,100180 
\citep[$2.0\pm0.4$~Gyr;][]{Barnes2007}, which has $P_{\rm rot}$=14~days 
and shows normal cycles on both sequences. The more evolved star 
HD\,143761 has $P_{\rm rot}$=17~days, and shows flat activity at $\log 
R'_{\rm HK}$=$-5.04$ for 25 years \citep{Baliunas1995}. The age from 
gryochronology implied by this rotation period is $2.5\pm0.4$~Gyr 
\citep{Barnes2007}, which agrees with the age of F-type stars observed by 
Kepler that have reached the critical Rossby number (blue dashed line) 
where the rotation period subsequently evolves much more slowly 
\citep{vanSaders2016}.

% FIGURE 3 --------------------------------------------------------------- 
%-------------------------------------------------------------------------

The transition across Ro\,$\sim$\,2 for G-type stars occurs at rotation 
periods comparable to the Sun ($P_{\rm rot}\sim$\,23--30~days). Before 
reaching this threshold, magnetic braking continues in these stars and 
their cycle periods evolve along the two sequences as their rotation 
slows. When they reach the critical Rossby number, the rotation rate 
changes much more slowly and we postulate that the cycle period responds 
to the magnetic transition. If we consider the evolutionary sequence 
defined by 18\,Sco \citep[4.1$\pm$0.5~Gyr;][]{Li2012, Mittag2016}, the Sun 
(4.6~Gyr), and $\alpha$\,Cen\,A \citep[5.4$\pm$0.3~Gyr;][]{Bazot2012}, the 
data suggest that a normal cycle on the lower sequence may grow longer 
across the transition (yellow dashed line). Eventually stars reach a low 
activity state like 16\,Cyg\,A\,\&\,B \citep[$P_{\rm rot}$$\sim$23.5~days 
at 7~Gyr;][]{Davies2015, Metcalfe2015}, where cyclic activity is no longer 
detected \citep{Hall2007a}. The Sun falls to the right of this 
evolutionary sequence because it is slightly less massive than the other 
stars (with a longer convective turnover time), so it does not reach the 
critical Rossby number until its rotation is a bit slower. Considering 
other sun-like stars, we propose that the solar cycle may be growing 
longer on stellar evolutionary timescales, and that the cycle might 
disappear sometime in the next 0.8-2.4~Gyr (between the ages of 
$\alpha$\,Cen and 16\,Cyg).

All of the slowest rotators with cycles ($P_{\rm rot}$$>$30~days) are 
K-type stars, which is now understandable---magnetic braking shuts down in 
more massive main-sequence stars before they reach these long rotation 
periods. Depending on the effective temperature, K-type stars reach the 
critical Rossby number at rotation periods longer than 35~days. The 
hottest cycling K-type star in our sample is HD\,219834A, and it appears 
to be well along the magnetic transition (red dashed line). All of the 
stars to the right of HD\,219834A are significantly cooler, so they have 
not yet reached the critical Rossby number \citep{Brandenburg2017}. The 
slightly hotter star HD\,182572 \citep[$P_{\rm 
rot}$$\sim$41~days;][]{Baliunas1996} appears to have already completed the 
magnetic transition like 16\,Cyg\,A\,\&\,B, showing flat activity at $\log 
R'_{\rm HK}$=$-5.10$ for 13 years \citep{Baliunas1995}.

Although $\alpha$\,Cen\,A appears in the same region of Figure~\ref{fig3} 
as several K-type stars, the broader evolutionary scenario suggests that 
the current cycle evolved from a shorter period on the lower sequence near 
18\,Sco. The expected cycle period on the upper sequence for G-type stars 
at the rotation period of $\alpha$\,Cen\,A is $\sim$35 years, much longer 
than the observed cycle \citep{Brandenburg2017}. In addition, there is no 
evidence of a shorter cycle in $\alpha$\,Cen\,A \citep[see][]{Ayres2014}, 
even though 18\,Sco shows a cycle on the lower sequence at essentially the 
same rotation period.

We can understand the evolution of the cycle toward longer periods during 
the magnetic transition by considering the variation of convective 
velocity with depth. The velocity is larger in the outer regions of the 
convection zone, and becomes progressively smaller in deeper layers 
\citep[e.g.,][]{Miesch2012}. Consequently, a star will initially exceed 
the critical Rossby number in the outer layers and the condition will only 
be met later in the deeper layers as the character of differential 
rotation shifts and the local rotation rate continues to slow 
\citep{Metcalfe2016}. If the lower sequence in Figure~\ref{fig3} 
represents a dynamo operating closer to the surface, while the upper 
sequence is the result of a dynamo driven in deeper layers 
\citep[e.g.,][]{Kapyla2016}, we speculate that the cycle period may grow 
longer as the magnetic transition proceeds and pushes the dynamo into 
deeper layers. The size of the convection zone might then set the overall 
timescale for completing the transition, when cycles disappear (or become 
extremely long) as in HD\,143761, 16\,Cyg\,A\,\&\,B and HD\,182572. 
However, other identifications of the underlying dynamos that are 
responsible for the two stellar sequences may not support this 
interpretation.

% SECTION 4 %%%%%%%%%%%%%%%%%%%%%%%%%%%%%%%%%%%%%%%%%%%%%%%%%%%%%%%%%%%%%%

\section{Discussion and Future Outlook}\label{sec4} 

Motivated by the recent discoveries of a rotational and magnetic 
transition in middle-aged stars \citep{vanSaders2016, Metcalfe2016}, we 
have identified the corresponding evolution of stellar activity cycles. A 
reinterpretation of previously published observations suggests that cycle 
periods grow longer along two sequences as magnetic braking slows the 
stellar rotation, but at a critical Rossby number (Ro$\sim$2) the surface 
rotation rate changes more slowly while the cycle gradually grows longer 
before disappearing. Evidence for this scenario exists for a range of 
spectral types, from the hotter F-type stars (HD\,100180, HD\,143761), to 
well-characterized solar analogs (18\,Sco, $\alpha$\,Cen\,A, 
16\,Cyg\,A\,\&\,B), to the cooler K-type stars (HD\,219834A, HD\,182572). 
The Sun appears to have aleady started this transition, and the solar 
cycle is expected to grow longer on stellar evolutionary timescales before 
disappearing sometime in the next 0.8-2.4~Gyr (between the ages of 
$\alpha$\,Cen and 16\,Cyg).

The greatest obstacle to understanding how the magnetic transition 
influences stellar activity cycles is the paucity of suitable 
observations. The bright sample of stars that were monitored for decades 
by the Mount Wilson survey have well-characterized long activity cycles 
and rotation periods, but their basic stellar properties are uncertain. In 
particular, the precise masses and ages that would allow us to identify 
evolutionary sequences are currently available for just a few stars 
\citep[e.g.\ asteroseismology of 18\,Sco and $\alpha$\,Cen\,A;][]{Li2012, 
Bazot2012}. This situation will soon improve, after the Transiting 
Exoplanet Survey Satellite \citep[TESS;][]{Ricker2014} yields 
asteroseismic data for bright stars across the sky during a two year 
mission (2018--2020). Although the time-series photometry will span only 
27 days for most TESS targets, this was sufficient to detect solar-like 
oscillations in hundreds of Kepler stars down to V$\sim$12 
\citep{Chaplin2011}. Similar detections are expected from TESS down to 
V$\sim$7 \citep{Campante2016}, particularly in F-type and hotter G-type 
stars with larger intrinsic oscillation amplitudes.

Although the basic stellar properties of the fainter Kepler stars are 
well-constrained from asteroseismology, chromospheric activity data have 
not been collected for long enough to detect stellar cycles. About a dozen 
stars in the \cite{vanSaders2016} sample were monitored in Ca\,HK several 
times per year during the Kepler mission \cite[2009--2013;][]{Karoff2013}. 
The cadence was insufficient to detect the shortest activity cycles, and 
the limited duration hindered the identification of longer cycles. So far, 
the only credible cycle in a Kepler target was detected using 
asteroseismic and photometric proxies of activity \citep{Salabert2016}, 
revealing a 1.5-year cycle on the lower sequence at $P_{\rm rot}\sim$\,11 
days. Most of the Kepler sample has already made the transition across 
Ro\,$\sim$\,2, so we might expect them to be ``flat activity'' stars like 
16\,Cyg\,A\,\&\,B \citep{Hall2007a}, but this remains to be seen. Future 
observations with the Las Cumbres Observatory (LCO) global telescope 
network promise to probe the onset and duration of the magnetic transition 
that drives the evolution and eventual disappearance of sun-like activity 
cycles.

\begin{acks} 
We would like to thank Axel Brandenburg, Ricky Egeland, Ed Guinan, Jeff 
Hall, Phil Judge, Savita Mathur and Benjamin Shappee for helpful 
discussions. This work was supported in part by NASA grants NNX15AF13G and 
NNX16AB97G, and by the ``Non-profit Adopt a Star'' program administered by 
White Dwarf Research Corporation. 
\end{acks}

\bibliographystyle{spr-mp-sola}

\end{article} 

\end{document}